\documentstyle[prl,aps,epsfig]{revtex}
\begin{document}
\title{Observation of non-gaussian conductance fluctuations at low temperatures in Si:P(B) at the metal-insulator transition}
\author{Swastik Kar, A.K.Raychaudhuri and Arindam Ghosh$^\dagger$}
\address{Department of Physics, Indian Institute of Science, Bangalore 560012, INDIA}
\date{\today}
\maketitle

\begin{abstract}
We report investigations of conductance fluctuations (noise) in doped 
silicon at low temperatures (T$<20$K) as it is tuned through the 
metal-insulator transition (MIT). The scaled magnitude of noise, 
$\gamma_H$, increases with decrease in T following an approximate power 
law $\gamma_H \sim T^{-\beta}$. At low T, $\gamma_H$ diverges as $n/n_c$ 
crosses 1 from the metallic side. We find that the distribution function 
and second spectrum of the fluctuations show strong non-gaussian behavior 
below 20K as $n/n_c$ decreases through 1. In particular, the observed 
distribution function which is gaussian for $n/n_c >> 1$, develops a 
log-normal tail as the transition is approached from the metallic side 
and eventually it dominates in the critical region. 

\end{abstract}
\vspace*{0.5cm}

The Anderson-Mott metal-insulator transition (MIT) has been one of the most 
studied areas in condensed matter~\cite{scaling,revs}. However, even after 
several decades of work~\cite{revs}, there remain many unanswered questions. 
A fundamental issue of considerable theoretical interest is that of 
conductance fluctuations. The two main aspects of this issue are the 
divergence of noise near the critical region and the onset of non-gaussian 
fluctuations. The origin of conductance fluctuation in the weak localization 
regime is dominated mainly by the Universal Conductance fluctuations(UCF), 
and is distinctly different from that in the strongly localized insulating 
regime ($\sigma_{T\to0}=0$) which show activated conductance. There are some 
experimental investigations of the magnitude of conductance fluctuations in 
thin films of $In_{x}O_{y}$~\cite{In02} at room temperature and in Si 
inversion layers~\cite{bogdan} at low temperatures which showed a divergence 
near the MIT. However, no experimental investigations have looked into the 
question of non-gaussian fluctuations at the MIT. In this report we present 
the results of the first experimental investigations of non-gaussian 
conductance fluctuations($f<10Hz$) near the MIT at low temperatures. We have 
investigated the above aspects of conductance fluctuations in bulk single 
crystals of doped Si as it is driven through the Anderson-Mott transition. 
Specifically, we find that the fluctuations, which have a gaussian 
distribution in the metallic regime becomes strongly non-gaussian with a 
log-normal distribution at the transition region. This also depends strongly 
on temperature. We enumerate below the important results:

\begin{enumerate}

\item{In a 3-D system like doped Si ( Si:P(B)) the noise magnitude shows clear signatures of divergence as $n/n_c$ decreases through 1 and $T\to0$.}

\item{As  $n/n_c \to1$ and the system crosses over to the insulating regime, the conductance  fluctuations becomes strongly non-gaussian as evident both from the second spectra (defined later) and Probability Distribution Function (PDF).}

\item{The PDF as well as the second spectra also depend on the temperature and both of them becomes more non-gaussian as T is decreased.}
\end{enumerate}

The samples (P doped and in some cases B compensated single crystals of Si) were taken from the same batch that were used earlier for transport studies near the MIT ~\cite{samples} with a range of doping levels $n$ from 3.3$\times10^{18}/cc$ to $1\times10^{19}/cc$, covering both sides of the critical doping level ($n_c$) for MIT (see table, figure 1). They were initially characterised by conductivity ($\sigma$) measurements down to $T=0.1K$. Figure 1 shows the variation of $\sigma$ as a function of  $T^{1/2}$. When fitted to the simple relation $\sigma$ = $\sigma_0$ + m$T^{1/2}$, samples with $\sigma_0>0$ are taken as metallic. In the samples with $n/n_{c} \leq 1$   the conductivity for $T<10K$ follows Mott's variable range hopping (VRH) with the relation $\sigma(T)=\sigma_{c}exp(-(T_{0}/T)^{1/4})$. $T_{0} \to 0$ as $n \to n_{c}$ from the insulating side. 

Conductance fluctuations (noise) was measured as a function of T for $2K<T<300K$ using a five-probe ac technique ~\cite{scoff1}. The time series of voltage fluctuations were collected directly into a computer using a 16-bit A/D card and was digitally processed to obtain the power spectra $S_V(f)$. A typical power spectra is calculated from a time series of 1-5$\times10^5$ data points at fixed temperatures which was controlled to within $\pm$1mK. Typically the power spectra $S_{V}(f)$ $\sim$$1/f^{\alpha}$ ($\alpha \sim 1$) and it scales inversely as the sample volume. At low measurement bias $S_{V}(f)$ scales as $V^{2}$. However, beyond certain bias values, deviation from $V^2$ dependence was seen at lower temperatures, and will be discussed later on. 

In all the samples, the noise magnitude increases as the temperature is decreased below 100K. For $T>100K$ the noise increases with T again  and this  arises from more classical mechanisms ~\cite{pramana,Kar} which are outside the scope of this paper. Figure~2 shows a plot of $\frac{\langle \Delta G^2\rangle}{G^2}$ = $\int df.S_V(f)/V^2$ integrated over the experimental bandwidth for a representative insulating sample K139 ($n/n_{c}$=0.94) for T$<$10K.  When measured with low enough biasing field, $\frac{\langle\Delta G^2\rangle}{G^2}$ shows an approximate inverse power-law dependence $\sim A/T^{\beta}$ with temperature. The behavior presented in figure~2 is typical for all the samples on both sides of the MIT, right down to the lowest measured T.

At lower temperatures in all the samples, the measured noise is sensitive to the biasing field used for the measurement as can be seen in figure~2. It shows how the T dependence can change drastically when measured with different biases. As T decreases, the bias dependence  of  $\frac{\langle\Delta G^2\rangle}{G^2}$ becomes more severe. On increasing the measuring bias beyond  a characteristic electric field $E^*$(T) (see figure 2 inset), $\frac{\langle\Delta G^2\rangle}{G^2}$ decreases drastically as a function of the measuring bias following an inverse power-law. $E^*$(T) in all samples decreased with T. We find that sample heating can explain only a small part of this noise suppression. The bias dependence noise has a strong impact on the measured temperature dependence of noise, and is clearly seen in all samples. Unless measured with a bias $E<E^*$, the noise tends to "flatten out" or saturate at lower temperatures, as seen in figure~2. The cause for this effect is currently under investigation and will not be elaborated any further in this paper. All the results presented henceforth have been taken with biasing fields $E<E^*$ such that $S_{V}(f) \propto V^{2}$ and has no bias dependence.

Figure 3 shows a plot of the Hooge parameter~\cite{DH} $\gamma_H$ (=$f.NS_V(f)/V^2$) as a function of $n/n_c$ for different temperatures for $f=1Hz$, $N$ being the total number of carriers in the measured volume. For $n/n_{c}>1$ (metallic regime), $\gamma_H$ is comparatively low. As the doping level decreases and $n/n_{c}\to 1$ from the metallic side, $\gamma_H$ increases slowly. For $n/n_{c}\approx$ 1, $\gamma_H$ starts increasing rapidly and at the MIT, changes drastically over several orders of magnitude as it goes over to the insulating side. On the insulating side, the observed $\gamma_H$ depends strongly  on the temperature (see inset 3a). At higher temperatures ($T \geq 6K$), the value of $\gamma_H$ passes through a sharp peak at a value of $n/n_c$ very close to but less than 1. It then drops sharply in the insulating side~\cite{refer1}. Interestingly this is roughly the same temperature ($T \approx 10K$) at which the resistivity starts to become activated and follows the VRH law. 

The inset (b) of figure~3 shows the variation of $\beta$ as a function of the doping level in the system. $\beta$ also has a strong n dependence close to the MIT. At the highest doping levels, $\beta \approx 0.5$, slowly increasing as the concentration is decreased and shows a rather sharp increase while it passes through the MIT to reach a value of 6 in the sample with the lowest doping level measured. The varitaion of $\beta$ in the metallic regime as the MIT is approached has been discussed elsewhere~\cite{ghosh0}.

On the weakly localized  metallic side ($n/n_{c} \geq 2$), we have shown earlier~\cite{ghosh1,ghosh2} that the predominant contribution to conductance fluctuations arise from the mechanism of UCF~\cite{UCF1}. In this region, the measured noise was found to be mostly gaussian (both PDF and the second spectra). However, there is  strong departure from such a gaussian behavior as  $n$ is reduced through $n_{c}$. A build up of non-gaussian  fluctuations as $n/n_c$ is reduced through 1  at low temperaure ($T=4.2K$)  can be seen from the probability distribution function (PDF). This is obtained from the time series and is the probability of occurence $P(\Delta V)$ of the voltage jump of magnitude $\Delta V$ when the conductivity fluctuates in a current biased sample. In all the samples (spanning different regions of $n/n_c$) the PDF calculated from the time series of the conductance fluctuations can be expressed as a sum of a gaussian and a log-normal distribution: $P(\Delta V) = A_{g}e^{-1/2(\frac{\Delta V}{\sigma_g})^{2}} + \frac{A_{ln}}{\Delta V} e^{-1/2(\frac{ln(\Delta V)-M}{\sigma_{ln}})^{2}}$.  $A_{g}$ and $A_{ln}$ are the relative weights of the two contributions. Figure~4a and b show the PDF of 3 representative samples  at 4.2K (in a liquid He bath). For the  metallic sample (K348) the distrubution is completely gaussian with $A_{ln}$ =0. However, close to the  the transition for the sample K242 one can see a gaussian distribution with a log-normal tail. For the insulating sample K240, there is no gaussian component ($A_{g} =0$) and the PDF is a log-normal distribution. The variation in the relative weights of the two contributions, as a function of $n/n_{c}$ , can be seen from the integrated area under the two distributions ($\Delta_{G}$  for the gaussian part and $\Delta_{LN}$ for the log-normal part) normalized by the total area under the PDF ($\Delta_{total}$) in figures~4(d)and (e). In figure~4(d) it can be seen that at $T=4.2K$, $\Delta_{G}$/$\Delta_{total}=1$ away from the transition but close to the t
ransition $\Delta_{G}$/$\Delta_{total}=0$. A complimentary behavior is seen for $\Delta_{LN}$/$\Delta_{total}$ . Figure~4(e) (done for sample K240) shows that as T is reduced from the high temperature the gaussian component is reduced while the log-normal component takes over. 

Another important measure of non-gaussianity is the second spectra, often referred to as the "noise in noise". If the fluctuations are uncorrelated, then the second spectrum~\cite{weissman} defined as $S^{(2)}(f)$ = $\langle(\Delta S^{(1)}(f))^2\rangle_{time}$ should have a frequency independent "white" frequency spectrum. $S^{(2)}(f)$ is a measure of higher order cumulants. A departure of $S^{(2)}(f)$ from a white spectrum will be an indication of  non-gaussian spectra and finite higher order cumulants. Figure~5a shows the second spectrum of the samples at T=4.2K. The spectra ($f<1Hz$)  fits the relation $S^{(2)}(f) \propto f^{-p}$.  As a function of $n/n_{c}$, $p$ changes (see inset, figure 5a) from a small value of $\sim 0.5$  for $n/n_{c} > 1$ (for white spectrum, $p=0$)to $\sim 1.5$ as the insulating state is reached, changing rapidly at $n=n_{c}$. Figure~5b shows how $p$ varies from a small value $\sim 0.5$ to a large value$\sim 1.3$ in the insulating sample K240 as the temperature is lowered. These observations clearly go to show that as the MIT is approached the dynamics that gives rise to the conductance fluctuation becomes increasingly non-gaussian and this is more so as the temperature is lowered. 

The divergence of the conductance fluctuation at the MIT has been predicted in some models. Since the conductance a variable with large random fluctuations, it has an important impact on the complete PDF. Some models~\cite{rossum} propose that in the large conductance limit (metallic side), the relative fluctuations are smaller and hence the higher cumulants of the fluctuations are negligible. This leads to a roughly gaussian PDF. With decrease in conductance, the relative fluctuations are larger and the higher cumulants start contributing. On the insulating side, this shows up as a log-normal PDF. Since the conductance decreases both as $n/n_c$ decreases through 1 as well as T decreases on the insulating side, our results are in excellent agreement with this model. However, there is one more important aspect in the non-gaussian fluctuation. The departure of the second spectra from a frequency independent white spectrum at low temperatures in samples with $n/n_{c}\leq 1$ clearly indicates that the non-gaussian fluctuation has a correlated dynamics associated with it. We have no clear interpretation as to what gives rise to it but we suggest that this may be an indication of a glass like freezing similar to what has been seen in spin glasses~\cite{weissman}. Recent theories~\cite{kirk1} have suggested that in the MIT region there is a glass like freezing of the charge system. Analysis of conductivity data close to the MIT using various scaling forms has shown that a comprehensive scaling description could be obtained if a glass-like transition is invoked at the MIT.  Recent noise experiments on 2-D electron system in Si inversion layers~\cite{bogdan} has shown that close to the MIT there is an appearance of large low-frequency dynamics which makes the noise blow up near the transition. This has been interpreted as a signature of freezing into electron glass. The divergence of noise at the MIT is unlikely to arise from the usual dynamics of atomic defect relaxations ~\cite{weissman,DH,UCF1} as the strongest coupl
ing mechanism that can produce conductivity fluctuation from atomic dynamics is UCF which itself saturates at a variance of $e^{2}/h$ (Mechanisms like UCF have limits to the fluctuations they can produce, i.e. when UCF is saturated). The large $\gamma_{H}$ seen in the insulating samples need much stronger sources of fluctuations. As the system goes through the MIT the dopant concentration is reduced and there is no reason why the defect dynamics should become more correlated at the MIT (for a $n$ $\sim 3\times 10^{18} cm^{-3}$ the average defect separation is $\approx 10nm$). In view of this it is most likely that the fluctuation that we see close to the MIT at low temperatures is electronic in origin and is a manifestation of the slowing down of the time scale of charge fluctuations as the MIT is approached. On approach to the MIT the single particle density of states can show both temporal and spatial fluctuations and if the dynamics is glassy then the relaxation time becomes an exponential function of the correlation length~\cite{kirk1}. We propose that the large rise in the low frequency noise at the approach to transition is a manifestation of low frequency dynamics of the fluctuation in local density of states which are strongly correlated by coulomb interaction in the absence of strong screening.

The authors wish to thank CSIR, Govt. of India for a fellowship (S.K) and a sponsored project (A.K.R).

\vspace*{0.1cm}
$^\dagger$ {\small Present address: Cavendish Laboratory, University of Cambridge, Madingley Road, Cambridge, UK.}

\centerline{\large\bf Figure Captions}
{\small
\vspace{10pt}
\noindent Figure 1: Conductivity($\sigma$) vs. temperature of the samples Si:P(B). $\sigma$ of the 3 most metallic samples PS24,PS41 and D150 have been scaled down by a factor of 5 for convenience. The table shows the relevent sample properties. 

\vspace{10pt}
\noindent Figure 2: $\frac{\langle\Delta G^2\rangle}{G^2}$ vs. temperature in the insulating sample K139 measured at different biasing voltages. When measured with the lowest bias, the noise scales as $T^{-\beta}$ for all the samples (for K139, $\beta =6$). The inset shows how $\langle\Delta G^2\rangle$ deviates from a $G^2$ dependence on application of larger measuring biases.

\vspace{10pt}
\noindent Figure 3: The scaled noise $\gamma_H$ for different values of $n/n_c$. Inset (a) shows the same graph, magnified around $n/n_c=1$. Inset (b) shows the $n/n_c$ dependence of $\beta$. 

\vspace{10pt}
\noindent Figure 4: 4(a),(b) and (c) shows the PDFs at 4.2K for K348, K242 and K240. 4(d) shows the relative contribution of gaussian and lognormal parts (see text) of the PDF as a function of n. 4(e) shows the same as a function of T in sample K240.

\vspace{10pt}
\noindent Figure 5: 5(a) Second spectrum $S^{(2)}(f)$ at 4.2K (see text). The graphs  have been shifted vertically  for visual clarity. The inset shows the sharp growth of frequency dependence at the critical concentration $n=n_c$ at 4.2K. 5(b) T dependence of the exponent $p$ of the second spectra in K 240. 
}

\begin{figure}[h]
\begin{center}

\epsfig{figure=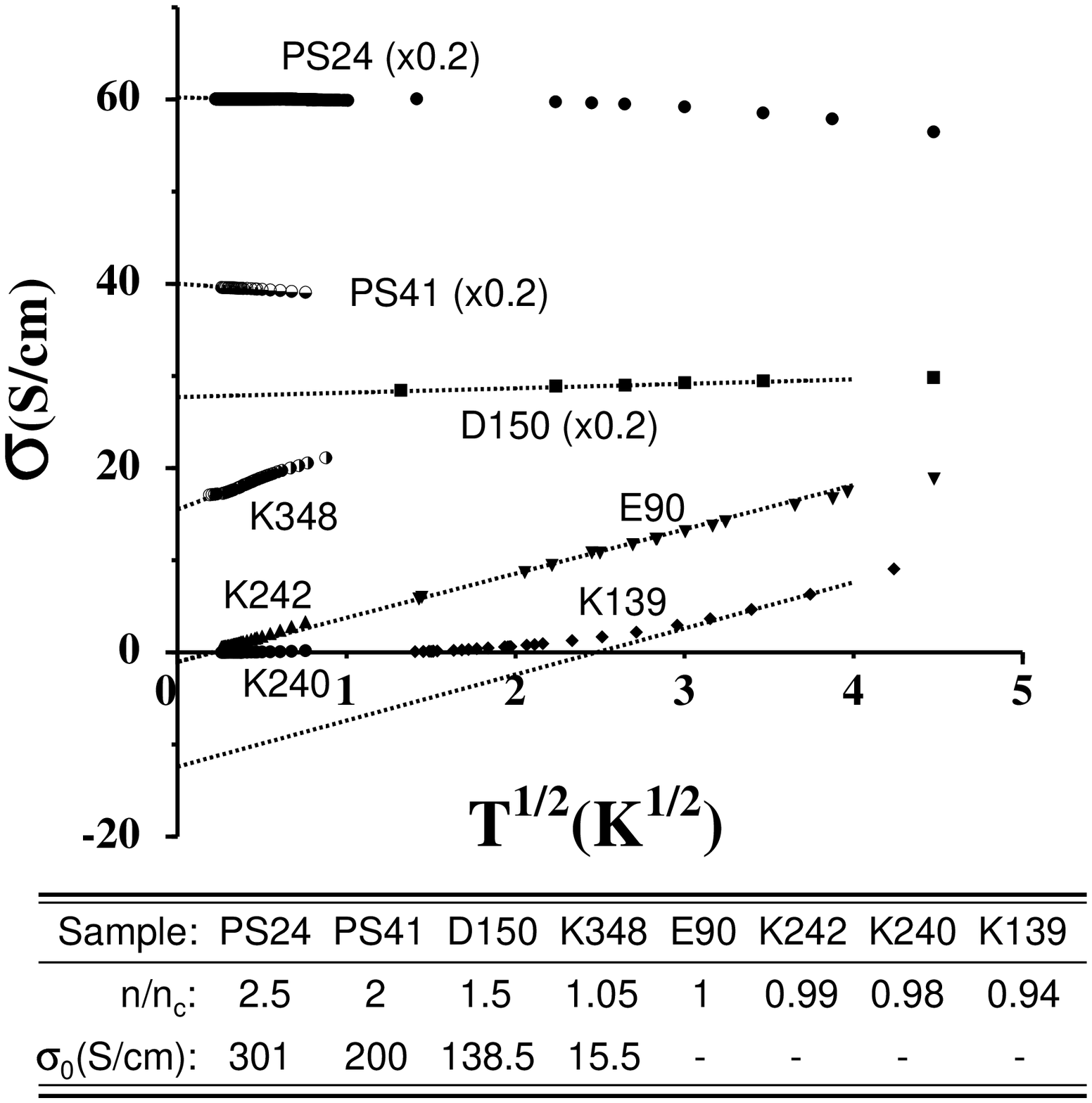,width=15cm,height=15cm,clip=}
\caption{}

\epsfig{figure=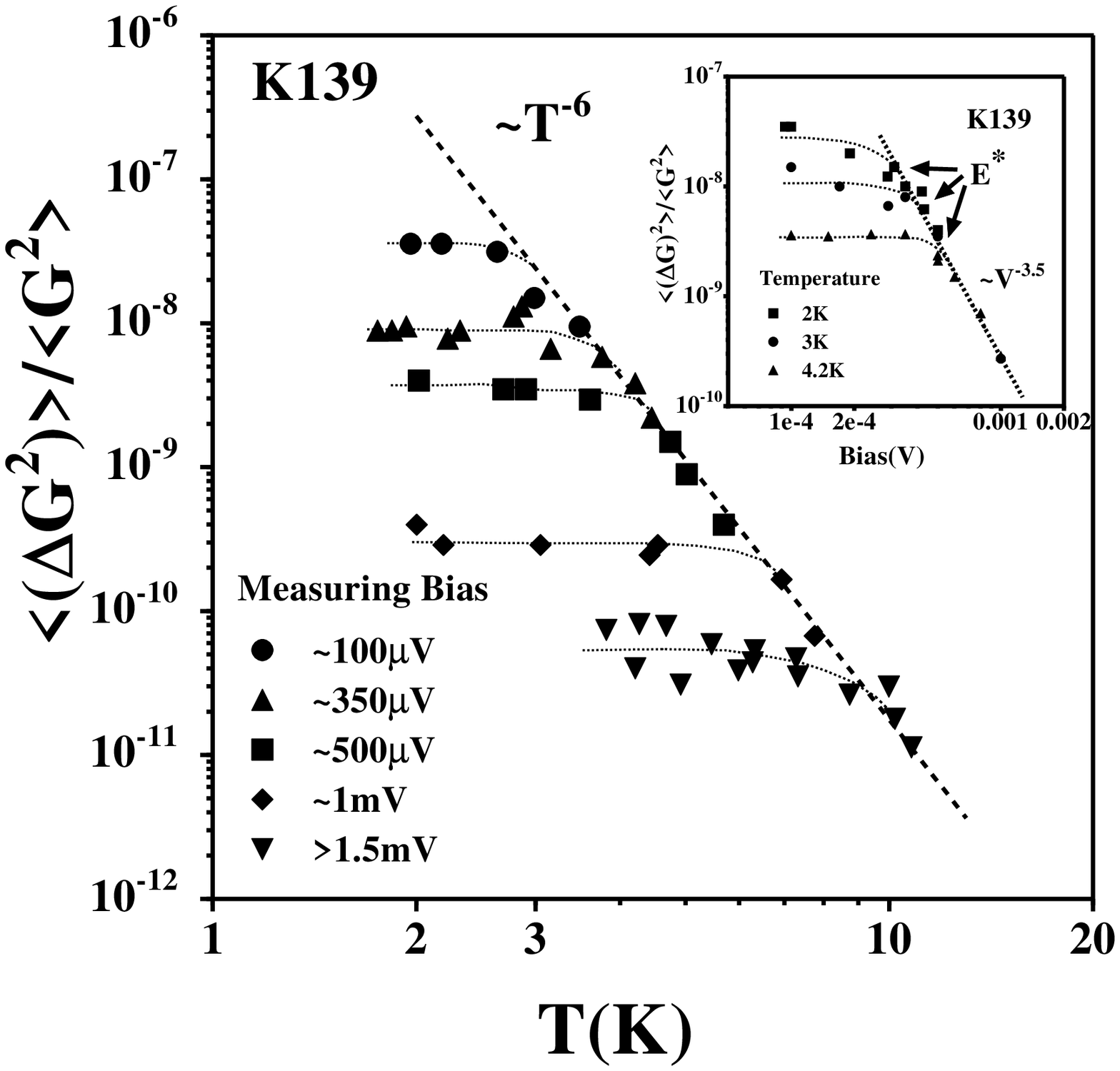,width=15cm,height=15cm,clip=}
\caption{}

\epsfig{figure=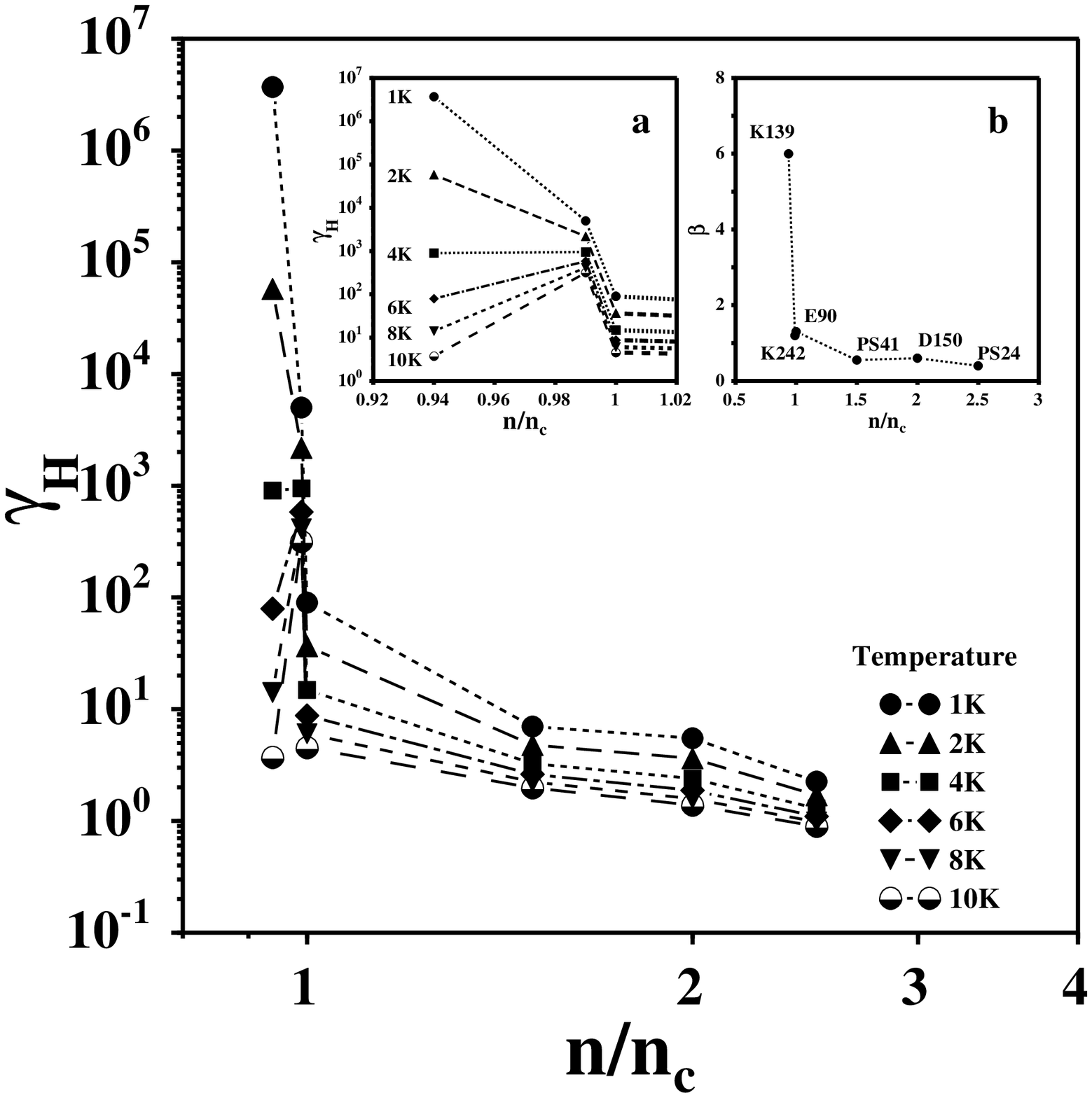,width=15cm,height=15cm,clip=}
\caption{}

\epsfig{figure=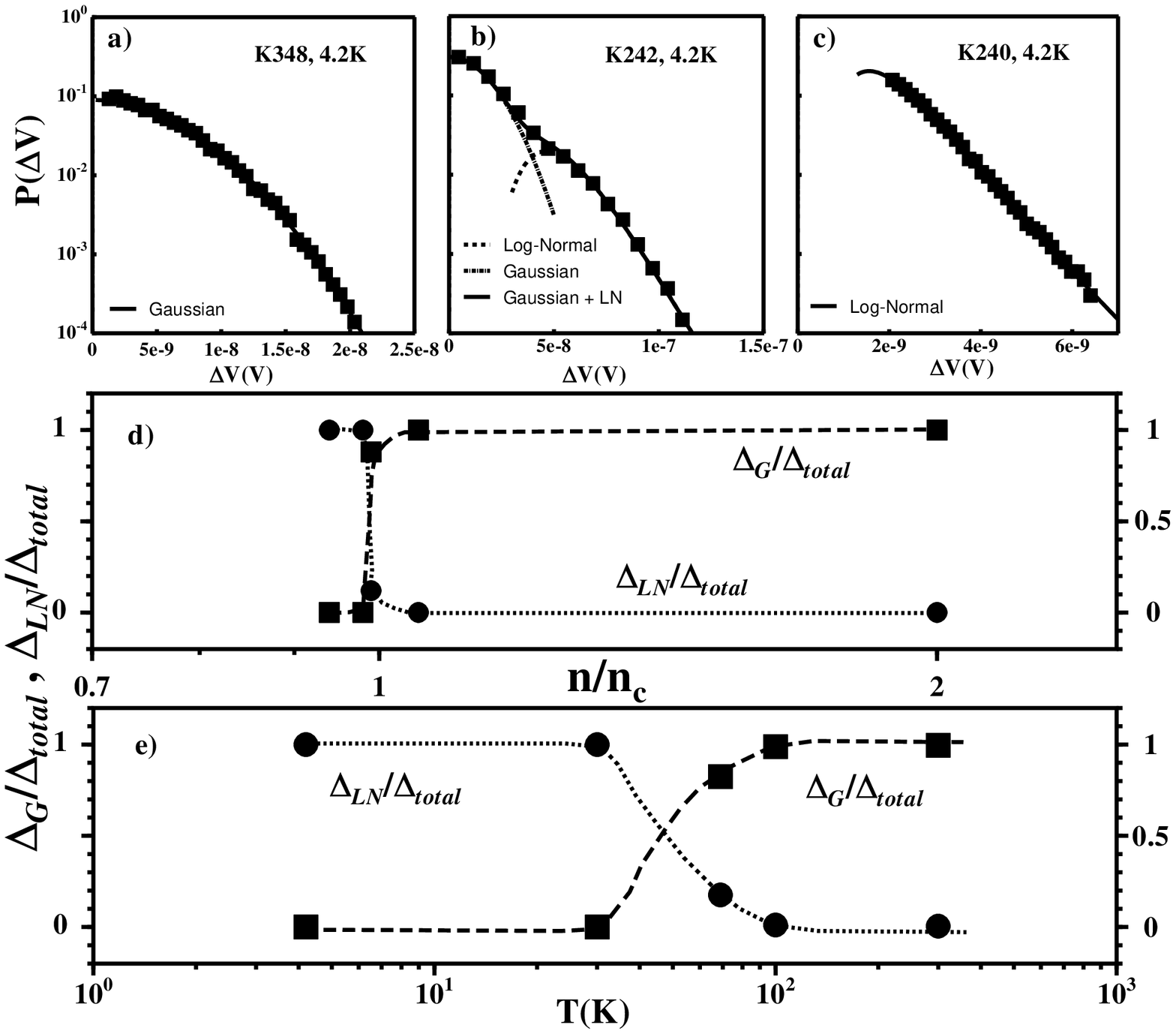,width=15cm,height=15cm,clip=}
\caption{}

\epsfig{figure=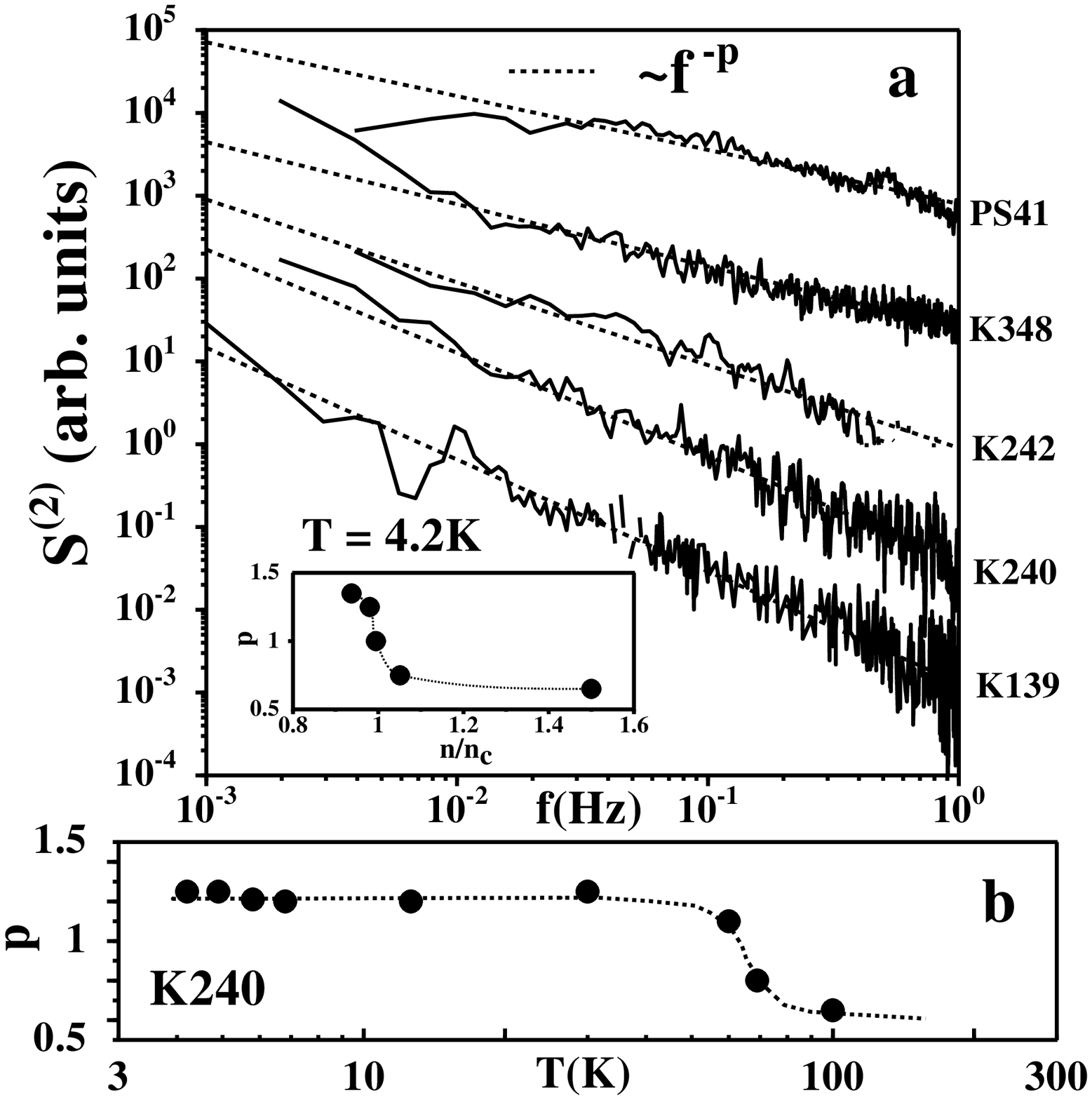,width=15cm,height=15cm,clip=}
\caption{}

\end{center}
\end{figure}

\end{document}